\documentclass[12pt]{llncs}
\usepackage{url}
\usepackage{graphicx}
\usepackage{color}

\newcommand{\Is}{\ensuremath{\mathbin{:=}}}

\newcommand{\set}[1]{\left\{ #1\right\}}
\newcommand{\floor}[1]{\left\lfloor #1\right\rfloor}

\newcommand{\RRem}[1]   {\`{\bf //\hspace{0.5mm}~}{\rm#1}}

\newcommand{\petaedge}[1]{#1}
\newcommand{\nopetaedge}[1]{}

\newcommand{\mytitle}{Scalable Generation of Scale-free Graphs}
\definecolor{mygray}{rgb}{0.5,0.5,0.5}
\definecolor{myred}{rgb}{1,0.5,0.5}
\definecolor{myblue}{rgb}{0.5,0.5,1}
\definecolor{mygreen}{rgb}{0.4,0.8,0.4}

\begin{document} 


\title{\mytitle}
\author{Peter Sanders, Christian Schulz}
\institute{Karlsruhe Institute of Technology (KIT), 76128 Karlsruhe, Germany\\ 
\email{\{sanders,christian.schulz\}@kit.edu}}
\date{today}
\pagestyle{plain}

\maketitle
\begin{abstract}
We explain how massive instances of scale-free graphs following the
Barabasi-Albert model can be generated very quickly in an
embarrassingly parallel way. This makes this popular model available
for studying big data graph problems. As a demonstration, we generated
a Petaedge graph in less than an hour.
\end{abstract}
\section{Introduction}
Scale-free graphs with a power-law degree distribution seem to be
ubiquitous in complex network analysis. In order to study such
networks and the algorithms to analyze them, we need simple models for
generating complex networks with user-definable parameters. Barabasi and Albert \cite{BA} define the model that is
perhaps most widely used because of its simplicity and intuitive
definition: We start with an arbitrary seed network consisting of
nodes $0..n_0-1$ ($a..b$ is used as a shorthand for $\set{a,\ldots,b}$
here).  Nodes $i\in n_0..n-1$ are added one at a time. They randomly
connect to $d$ neighbors using \emph{preferential attachment}, i.e.,
the probability to connect to node $j\leq i$ is chosen proportionally
to the degree of $j$. The seed graph, $n_0$, $d$, and $n$ are
parameters defining the graph family. 

With the recent interest in big data, the inherently sequential
definition of these graphs has become a problem however, and other,
less natural models have been considered for generating
very large networks, e.g., for the well known Graph500 benchmark
\cite{Graph500}.

\section{Our Algorithm}
Our starting point is the fast, simple, and elegant sequential
algorithm by Batagelj and Brandes \cite{Brandes}.  For simplicity of
exposition, we first consider the most simple situation with an empty
seed graph and where self-loops and parallel edges are allowed. See
Section~\ref{s:generalizations} for generalizations.

We use an empty seed graph ($n_0=0$).  A generalization
only requires a number of straight forward index transformations.
Batagelj and Brandes' algorithm generates one edge at a time and writes it into an
edge array $E[0..2dn-1]$ where positions $2i$ and $2i+1$ store the node
IDs of the end points of edge~$i$.  We have $E[2i]=\floor{i/d}$.  The central
observation is that one gets the right probability distribution for the other end point by
uniformly sampling edges rather than sampling dynamically weighted
nodes, i.e., $E[2i+1]$ is simply set to $E[x]$ where $x$ is chosen
uniformly and (pseudo)randomly from $0..2i$.

The idea behind the parallel algorithm is very simple -- compute edge
$i$ independently of all other edges and without even accessing the
array $E$. On the first glance, this sounds paradoxical because there
\emph{are} dependencies and accessing $E$ is the whole point behind
Batagelj and Brandes' algorithm.  This paradox is resolved by the idea to 
recompute any entry of $E$ that is needed for edge~$i$. However, this
solution raises two concerns. First, doesn't recomputation increase
the amount of work in an inacceptable way? Second, how do you
reproduce random behavior? 

The first concern is resolved by observing that we have a 50 \% chance
of looking at an even position which is easy to reproduce. In the other
50 \% of the cases, we have to look at
further positions of $E$. Overall, the expected number of positions of
$E$ considered is bounded by $\sum_{i\geq 0}2^{-i}=2$ -- compared to Batagelj and Brandes' algorithm, we
compute around twice as many random numbers but in exchange save most
of the expensive memory accesses.

What saves the situation with respect to reproducing random behavior
is that practical computer programs usually do not use true randomness
but only pseudorandomness, i.e., if you know the state of a
pseudorandom number generator, you can reproduce its results
deterministically. This is a nontrivial issue in a parallel setting
but becomes easy using another trick. We use a hash function that maps
the array position to a pseudorandom number. For a graph generator
this approach has the additional benefit that the graph only depends
on the hash function but not on the number of processors used to
compute the graph. The pseudo-code below summarizes the resulting
algorithm where $h(r)$ is a hash function mapping an integer $r$ 
to a random number in $0..r-1$.

\begin{figure}[t]

\begin{tabbing}%
\hspace{1.5em}\=\hspace{1.5em}\=\hspace{1.5em}\=\hspace{1.5em}\=\hspace{1.5em}\=\hspace{1.5em}\=\hspace{1.5em}\=%
\hspace{1.5em}\=\hspace{1.5em}\=\hspace{1.5em}\=\hspace{1.5em}\=\hspace{1.5em}\=%
\kill

{\bf function} generateEdge$(i)$\RRem{generate $i$-th edge}\+\\
  $r\Is 2i+1$\\
  {\bf repeat}
    $r\Is h(r)$
  {\bf until} $r$ is even\\
  {\bf return} $(\floor{i/d},\floor{r/2d})$
\end{tabbing}

\caption{Pseudocode of generateEdge.}
\end{figure}
Note that this simple setting allows us a number of interesting
approaches. Using dynamic load balancing, handing out batches of edge
IDs, we can use cheap heterogeneous cloud resources. We do not even
have to store the graph if the analysis uses a distributed streaming
algorithm that processes the edges immediately.

\section{Experiments}

\begin{figure}[t!]
\centering
\nopetaedge{\includegraphics[width=\textwidth]{degree_plot_n12_d100.png}}
\petaedge{\includegraphics[width=\textwidth]{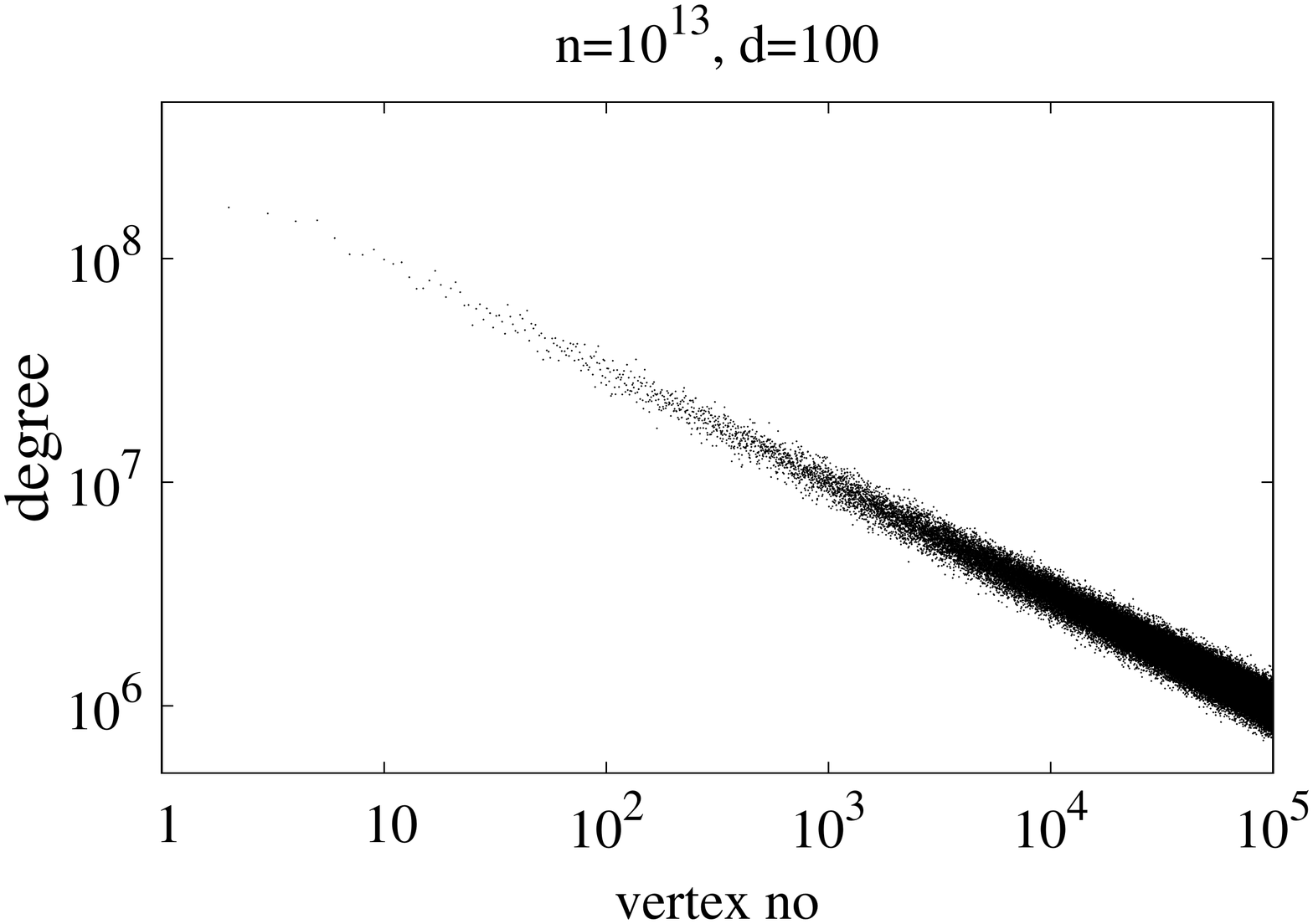}}
\caption{The degrees of the first $100K$ nodes of a graph with \nopetaedge{$n=10^{12}$}\petaedge{$n=10^{13}$} nodes and $m=100n$ edges computed on 16\,384 cores of the SuperMUC supercomputer. }
\label{fig:degrees}
\end{figure}

 As a simple example,
we have implemented an algorithm finding the degrees of the first $100K$ 
nodes of a graph with \nopetaedge{$n=10^{12}$ nodes and $m=100n$}\petaedge{$n=10^{13}$ nodes and $m=100n$} edges on 16\,384 cores of the 
SuperMUC supercomputer\footnote{The authors gratefully acknowledge the Gauss Centre for Supercomputing e.V. (\url{www.gauss-centre.eu}) for funding this project by providing computing time on the GCS Supercomputer SuperMUC at Leibniz Supercomputing Centre (LRZ, \url{www.lrz.de}).} (see Figure~\ref{fig:degrees}).
 As a hash
 function we use the CRC32 instruction available since SSE 4.2 twice with random seeds to
 obtain 64 bits of pseudo-random data (see Figure~\ref{fig:hashfun}). Preliminary experiments indicated that this is slightly faster than using a simple hash function.
In comparison, Batagelj and Brandes algorithm is about 60 \% faster than our
algorithm on a single core but slower than any parallel
run.  We are about 16 times faster than the parallel RMAT generator~\cite{Graph500} for a graph with $50\cdot 10^9$ edges.  Alam et
al.~\cite{Alam} report 123 s for generating a graph with $50\cdot 10^9$
edges on 768 processors. They use an algorithm explicitly tracing dependencies which leads to massive amounts of fine grained communication. We are about 36 times faster on a machine
with slower cores and using 64 bit node IDs rather than 32 bits.
Meyer and Penschuk \cite{MeyPen16} can generate a graph with the same
parameters on a single node with 16 cores and a GPU using an ingenious
external memory algorithm and 6 SSDs in 489 s. Our streaming generator using
just the 16 cores needs about 154 s.  
For smaller graphs, using 32 bits instead of 64 bits everywhere in our code saves about 65\% of running time. 
The largest graph we have
generated is 20\,000 times larger than the largest Barabasi-Albert graph we have seen reported.


\section{Generalizations}\label{s:generalizations}
A seed graph with $n_0$ nodes and $m_0$ edges is incorporated by
stopping the repeat loop in the edge generation function when $r<m_0$.  For $r<m_0$ the node ID is
precomputed. Otherwise, we compute $\lfloor(r-m_0)/d\rfloor+n_0$.
Self-loops can be avoided by initializing $r$ to the first edge of the
current node.  To avoid parallel edges, we store the target node IDs
generated for a node in a hashtable of size $\leq d$ and reject
duplicate target nodes.
Now consider a situation with individual degrees $d_i$ for each node.
If about $\sum_i\log d_i$ bits fit into each processor, we can
replicate a succinct representation of a sparse bit vector \cite{GogPet14}
with a one bit for each first edge of a node. Then the target node of
edge $r$ can be computed in constant time using the operation
rank$(r)$. For larger networks, we can defer the computation of node
IDs and first only compute edge IDs%
\footnote{We would like to thank Ulrik Brandes for pointing this
  out.}.  At the end, we can sort these edge IDs and merge the result
with the prefix sum of the node degree sequence to obtain the node
IDs. Any parallel sorting and merging algorithm can be used for this
purpose.
\begin{figure}[t]

\begin{tabbing}%
\hspace{1.5em}\=\hspace{1.5em}\=\hspace{1.5em}\=\hspace{1.5em}\=\hspace{1.5em}\=\hspace{1.5em}\=\hspace{1.5em}\=%
\hspace{1.5em}\=\hspace{1.5em}\=\hspace{1.5em}\=\hspace{1.5em}\=\hspace{1.5em}\=%
\kill
{\bf function} hash\_crc(x) \{\+ \\
        hash  $=$ \_mm\_crc32\_u64(0, x);  \\
        hash        $=$ hash $\ll$ 32;  \\
        hash        $+\!\!=$ \_mm\_crc32\_u64(1, x);  \\
        {\bf return} hash \% x; \-\\
\} \\
\ \\
        hashMultiplier=3141592653589793238LL; \\ 

{\bf function} h\_simple(x) \{ \+
{\bf return} (x*hashMultiplier)*(1.0/0xffffffffffffffffLL) * x; 
\}
\end{tabbing}

\caption{Hash functions used in our implementations.}
\label{fig:hashfun}
\end{figure}

\vfill\pagebreak
\bibliographystyle{plain}
\bibliography{BAgen}

\begin{thebibliography}{1}

\bibitem{Graph500}
Graph 500 Benchmark {\url{www.graph500.org/}}.

\bibitem{Alam}
Maksudul Alam, Maleq Khan, and Madhav~V Marathe.
\newblock Distributed-memory parallel algorithms for generating massive
  scale-free networks using preferential attachment model.
\newblock In {\em Proc. of the Int. Conference on High Performance Computing,
  Networking, Storage and Analysis}, page~91. ACM, 2013.

\bibitem{BA}
Albert-Laszlo Barabasi and Reka Albert.
\newblock Emergence of scaling in random networks.
\newblock {\em Science}, 286(5439):509--512, 1999.

\bibitem{Brandes}
Vladimir Batagelj and Ulrik Brandes.
\newblock Efficient generation of large random networks.
\newblock {\em Physical Review E}, 71(3):036113, 2005.

\bibitem{GogPet14}
Simon Gog and Matthias Petri.
\newblock Optimized succinct data structures for massive data.
\newblock {\em Software: Practice and Experience}, 44(11):1287--1314, 2014.

\bibitem{MeyPen16}
Ulrich Meyer and Manuel Penschuk.
\newblock Generating massive scale free networks under resource constraints.
\newblock In {\em Meeting on Algorithm Engineering \& Experiments (Alenex)},
  pages 39--52. {SIAM}, 2016.

\end{thebibliography}

\end{document}